\documentstyle[aps,pre,twocolumn,psfig,floats]{revtex}
\begin{document}
\wideabs{
\title{Compaction of Rods: Relaxation and Ordering in Vibrated,
Anisotropic Granular Material}
\draft
\author{Fernando X. Villarruel, Benjamin E. Lauderdale, Daniel
M. Mueth, and Heinrich M. Jaeger}
\address{The James Franck Institute and Department of Physics\\
	The University of Chicago\\
	5640 S. Ellis Ave., Chicago, IL 60637}
\maketitle

\bibliographystyle{prsty}

\begin{abstract}
We report on experiments to measure the temporal and spatial evolution
of packing arrangements of anisotropic, cylindrical granular material,
using high-resolution capacitive monitoring.  In these experiments,
the particle configurations start from an initially disordered,
low-packing-fraction state and under vertical vibrations evolve to a
dense, highly ordered, nematic state in which the long particle axes
align with the vertical tube walls.  We find that the orientational
ordering process is reflected in a characteristic, steep rise in the
local packing fraction.  At any given height inside the packing, the
ordering is initiated at the container walls and proceeds inward.  We
explore the evolution of the local as well as the height-averaged
packing fraction as a function of vibration parameters and compare our
results to relaxation experiments conducted on spherically shaped
granular materials.
\end{abstract}
 

%
}

\section{Introduction}
The packing of identical objects inside a given volume, from atoms to
large molecules and polymers to macroscopic particles, is an important
problem in many areas of science and technology.  For thermal systems,
there are well-defined optimal packing configurations, corresponding
to thermodynamic equilibrium.  In particular, at high packing
densities, these equilibrium configurations correspond to ordered,
often crystalline, states.  However, in many systems, such as glasses,
competing interactions between the individual constituents prevent a
thermodynamic equilibrium from being reached over experimentally
accessible time scales.  Furthermore, there are large classes of
nonthermal systems, including foams and granular materials such as
sand, rice, or pharmaceutical pills, for which ordinary temperature is
irrelevant and the usual concept of a corresponding, thermodynamic
equilibrium does not exist.  In these systems, local
temperature-driven fluctuations do not couple to particle motion in an
effective manner and do not allow for a full exploration of
configuration space.  At high packing densities, the most stable,
ordered configurations are therefore rarely reached and the packing
becomes trapped in metastable, disordered states.  Important questions
for these strongly non-equilibrium systems remain only partially
answered, such as how the metastable configurations respond to
perturbations, and how the packing fraction (i.e., the fraction of
volume occupied by the granular material) evolves over long times.

Macroscopic granular materials provide a model system for the
exploration of these issues~\cite{mehta94,degennes99,jaeger96}.  In a
three-dimensional (3-D) packing of monodisperse, rigid spheres held
together by gravity and frictional forces, there is a myriad of
metastable states, each corresponding to a different packing
configuration that satisfies mechanical equilibrium, yet with packing
fraction $\rho$ far less dense than the most stable, crystalline
configuration for which $\rho \approx 0.74$.  Configuration space can
be explored conveniently through the application of external
mechanical excitations, such as shaking or vibrating.  Starting from
an initial, low-packing-fraction configuration, the packing evolves
over time toward an asymptotic, higher-packing-fraction state with a
more compact particle arrangement.  For disordered 3-D packings of
equal-sized spheres, computer simulations, as well as experiments on
steel balls, have found~\cite{finney70,yen92} that the maximum final
packing fraction is set by the random close packing limit, $\rho_{\rm
rcp} \approx 0.64$.  Physically, this limit corresponds to amorphous
packing configurations that are fully frustrated by geometrical
constraints and unable to compact further.  We note in this context
that a (topological) effective temperature of the packing may be
defined in terms of the available free volume~\cite{shahinpoor80} or
compactivity~\cite{edwards89}.  Increases in volume fraction then
correspond to decreases in this temperature.  High densities around
$\rho_{\rm rcp}$ are only reachable via careful cycling of the excitation
intensity~\cite{nowak97a}, similar to thermal cycling during annealing
procedures.  For fixed excitation intensity, the packing will evolve
toward final configurations reflecting a balance between defect
creation and annihilation.  This suggests a second (dynamic) type of
effective temperature associated with the strength of the applied
forcing.  Recent experiments showed that the corresponding final
densities $\rho$ are approached logarithmically slow in
time~\cite{knight95,nowak98}.  In a number of theoretical models this
slow relaxation was explained as arising from geometrical frustration
due to excluded
volume~\cite{nowak98,boutreux97,caglioti97,gavrilov98,head98,linz98,bennaim98},
and analogies to glassy behavior were
drawn~\cite{barker93,coniglio96,nicodemi97,kolan99}.  Furthermore, it
was found that, at short times and low excitation levels, the system
can only explore a limited region of configuration space, resulting in
highly irreversible behavior and memory effects.  Only after
sufficiently long times and large excitation levels does one reach a
reversible, steady-state response~\cite{nowak97a,kolan99} in the sense
that the two effective temperatures track each other, i.e., the
packing fraction becomes a monotonic function of applied forcing.

Actual materials, and certainly macroscopic granulates, typically are
far from perfectly spherical and often elongated.  Under thermal
conditions, particle anisotropy is known to give rise to ordering in a
variety of systems, such as liquid crystals~\cite{chaikin95}, rod-like
colloidal virus particles~\cite{dogic97}, and certain
polymers~\cite{ciferri91}.  For example, a fluid of long, rigid rods
at high packing fraction will undergo a transition to an ordered
nematic phase in which the rod axes align along a common
direction~\cite{onsager49}.  In nonthermal systems, on the other hand,
almost all work to date has focused on spherical particles and the
effect of particle anisotropy on the stability and evolution of
packing configurations has been largely unexplored.  Recent
theoretical work by Baulin and Khokhlov~\cite{baulin99} on sedimenting
solutions of long rigid rods investigated the limit that external
(gravitational) forces far outweigh thermal fluctuations and predicts
an isotropic to nematic transition for increasing packing fraction.
The nature of the transition into the nematic state was studied by
Mounfield and Edwards~\cite{mounfield94} for a model of granular rods.
They concluded that an externally imposed (flow) field was required to
stabilize a discontinous, first-order-type phase transition into a
highly ordered nematic state; otherwise there would merely be a
cross-over during which the ordering increases continuously with
decreasing compactivity (or increasing packing fraction).  Buchalter
and Bradley~\cite{buchalter92,buchalter94} used Monte Carlo simulations
to study packings of rigid, prolate or oblate ellipsoids.  They found
that these systems form amorphous, glassy packings with long-range
orientational order.  Some limited experimental data on the compaction
of non-spherical, prolate granular materials under vibrations has been
published by Neuman~\cite{neumann53}, but, to the best of our
knowledge, no information on the degree of ordering or on the
asymptotically reached, 3-D packing configurations is available.  This
lack of systematic investigations is surprising, given the enormous
importance of prolate granular materials in a wide range of geological
and industrial processes.

Here we study 3-D packings comprised of prolate granular material:
millimeter-sized rigid cylinders (``rods'').  Applying discrete mechanical
excitations, or ``taps'', we let the system evolve from an initial,
low-packing-fraction to a final, high-packing-fraction state.  During this
relaxation process, we monitor the local packing fraction non-invasively
and correlate it with direct visual images of the packing configurations.
We are specifically interested in the question of how two competing factors
affect the packings' evolution as the packing fraction increases: on the
one hand, the tendency of randomly arranged rods to lock up in a disordered
state because of steric hindrance and friction, and on the other hand the
possibility, provided by both gravity and container walls, to align and
form ordered, nematic-type configurations.  Our results show that there are
characteristic stages to the evolution, corresponding to either process.
We find that, depending on the tapping history and intensity, highly
ordered final states are achievable, in contrast to sphere packings under
the same experimental conditions.

This paper is organized as follows.  In Section II we describe the
experimental set-up and procedure.  Results on the relaxation and alignment
behavior for a range of excitation intensities are shown in Section III and
discussed in Section IV.  Section V contains a summary of the findings and
conclusions.

\section{Experimental Set-Up and Procedure}
All experiments were performed on monodisperse nylon 6/6 rods of specific
density ($1.145\pm 0.005$)g/${\rm cm}^3$, each 1.8~mm in diameter and 7.0~mm in length.
Approximately 7200 of these rods were filled into a 1~m tall glass tube
(1.90~cm inner diameter) mounted vertically on a Bruel and Kjaer 4808
electromagnetic vibration exciter (Fig.~1a).  As in previous experiments on
spherical particles~\cite{nowak97a,knight95,nowak98,nowak97b}, vertical vibrations were applied in the
form of individual shaking events (``taps'') by driving the exciter with a
single cycle of a 30~Hz sine wave.  Successive taps were spaced by time
intervals sufficiently long (typically 0.5~s) to allow complete relaxation
of the system.  The vibration intensity was monitored using an
accelerometer.  In the following we parameterize the tapping intensity by
$\Gamma$, the ratio of the measured peak acceleration to the Earth's acceleration
${\rm g} = 9.81 {\rm m/s^2}$ (Fig.~1c).  The bottom of the tube contained an entry way for
dry nitrogen, which was used only in the preparation of the initial, low
packing fraction state of the sample.  Through the top of the glass tube
the system was placed under vacuum during runs.  A control tube similar to
the tube described above, filled with the same material but not vibrated,
was used to measure electronic drift.

\begin{figure}
\centerline{
\psfig{file=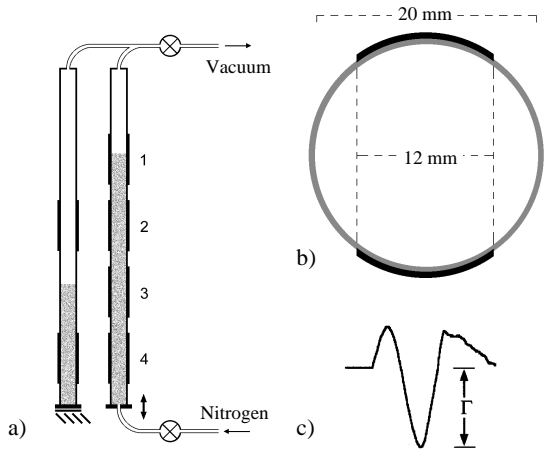,width=3in}}
\vspace{2ex}
\caption{
(a) Schematic of the experimental system used for capacitive
measurements of the packing fraction. The set-up consists of a
measurement column with four capacitors and a control column with two
capacitors for tracking electronic drift.  The initial packing state
is prepared by fluffing with dry nitrogen injected through the valve
shown at the bottom.  (b) Sketch of the measurement tube
cross-section.  Two capacitor electrodes are shown at the tube
perimeter.  The active region sensed by the capacitor lies between the
dashed lines. (c) Time trace of an individual, vertical ``tap'' as
measured by the accelerometer.  The peak acceleration, $\Gamma$, is
indicated.  }
\label{fig:apparatus}
\end{figure}

The evolution of the packing fraction between taps was monitored both
globally by recording the total filling height of material inside the
tube, and locally using a capacitive technique.  Mounted along the
outside of the tube were four capacitors made from pairs of copper
strips, each 1.25~cm wide and 17~cm in length.  Each of the capacitors
was sensitive to a measurement volume inside the tube defined by a
slab of cross-sectional area as indicated in  Fig. 1b (about 70\% of
the total cross-sectional area of the tube).  We ascertained that
there was little sensitivity to material placed outside this active
volume.  The capacitance was read by a capacitance bridge with 1fF
resolution.  The relation between packing fraction, $\rho$, and
capacitance, $C$, was found to be linear, $\rho = -3.47\times 10^{-1} +
2.35\times 10^{-3} C$, where $C$ is measured in fF(Fig.~2).  The data in
this figure contains data from all capacitors, with readings for the
empty ($\rho = 0$) tube and for a solid nylon rod occupying its total
volume ($\rho=1$), as well as data for intermediate packing fractions.
For the latter, the two methods employed, a) inserting a solid rod
partially into a capacitor or b) compacting a test sample of cylinders
with a known number of particles inside the measurement volume, both
yielded the same results, indicating that particle shape effects do
not significantly influence the packing fraction measurements.

\begin{figure}
\centerline{
\psfig{file=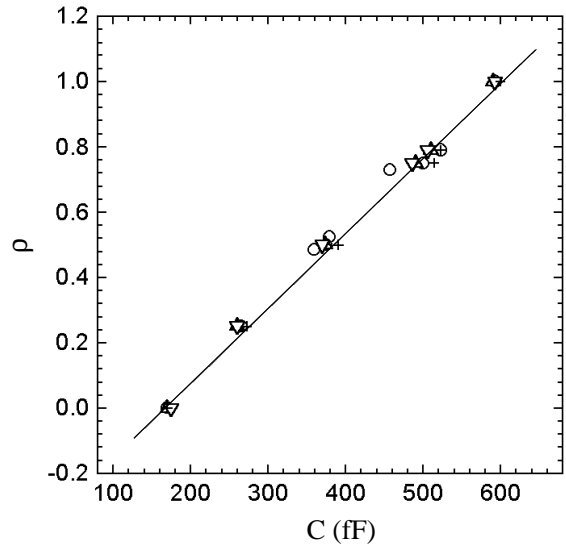,width=3in}}
\vspace{2ex}
\caption{
The capacitor calibration used to determine the packing fraction,
$\rho$, from the measured capacitance, $C$, at various heights in the
column. The circles, triangles, down triangles, and crosses are for
capacitors 1, 2, 3, and 4, respectively. The data is seen to be
well-fit by a linear relationship.  }
\label{fig:calibration}
\end{figure}

Because visual tracking of particles through the tube side walls was
limited to areas between capacitors, we used a separate set-up with a
tube of same diameter but shorter (21~cm) and without capacitors to
explore qualitatively the evolution of particle orientations.  This
was done by video-taping the material through the outside walls and
also from above.  In addition, careful layer-by-layer vacuuming
allowed us to map out the depth dependence.

For each run, the 1~m tall tube was filled with 143.1~g of material to
a height of ($82.3 \pm 0.3$)cm.  The material was then fluffed with
nitrogen to an initial filling height of ($90.6 \pm 0.5$)cm,
corresponding to a packing fraction $\rho = 0.49$.  We found this to
represent the least dense, reproducible initial packing state
attainable.  Both measurement and control tube were then placed under
vacuum to isolate them from environmental changes in the room during a
run.

Here we describe two sets of experiments.  In the first set, the
material was tapped 70,000 times at fixed acceleration and capacitance
readings were made after certain, fixed tap intervals.
Simultaneously, the average packing fraction of the column was
recorded by measuring the overall filling height with a ruler.  Prior
to each experimental run, the system was returned to the same
loose-packed initial state by removing all material, refilling and
fluffing with nitrogen.  This was done to remove all traces of
ordering from previous runs.  In the second set of experiments, we
explored the effect of tapping history on the packing fraction
evolution.  A fixed number, $\Delta t$, of taps were applied to the
system and the final packing fraction was recorded.  Without refilling
or fluffing the system, the acceleration was adjusted by
$\Delta\Gamma$ and the measurement process was continued, ramping
$\Gamma$ from 1.5 to 7.5 and back several times.  This is similar to
what would be done in a cyclic heating and cooling process, with
$\Delta\Gamma/\Delta t$  playing the role of an effective heating or
cooling rate.

\section{Results}
Figure~3 shows the evolution of the packing fraction, $\rho(t)$, for
different capacitor regions as a function of  tap number, $t$.  Data
for four different accelerations are shown.  Each curve is an average
of five independent runs and the error bars, for sake of clarity given
only for capacitor~2 in the $\Gamma = 4.5$ graph, represent the
general rms variation.  In order to be able to display the initial
packing fraction, $\rho(0)$, the tap number (or time) axis on all
plots has been incremented by one tap.

The evolution shown in Figure~3 exhibits three distinct stages which
we call the initial relaxation stage (up to about $10^3$ taps), the
vertical ordering stage (between roughly $10^3$ and $10^4$ taps), and the
final, steady-state regime.

\begin{figure}
\centerline{
\psfig{file=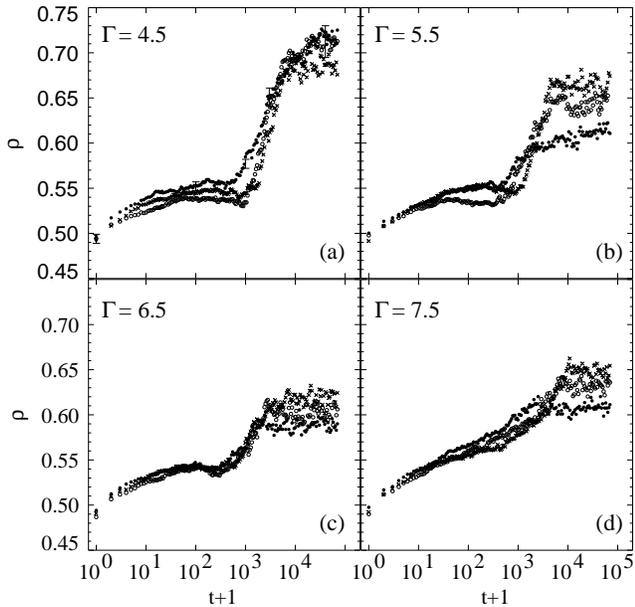,width=3.3in}}
\vspace{2ex}
\caption{
The evolution of the packing fraction, $\rho(t)$, as a function of tap
number, $t$, for several accelerations, $\Gamma$, as indicated. Each
plot contains traces for regions 2, 3, and 4 along the height of the
tube (Fig.~1a), shown by solid circles, open circles, and crosses
respectively (we excluded region 1 since it was partially emptied
during the compaction of the column).  Most pronounced for smaller
$\Gamma$, we observe three distinct regimes, corresponding to separate
physical processes as explained in the text.  }
\label{fig:relaxation}
\end{figure}

During the initial relaxation stage there is a quick increase in
$\rho$ during the first decade and then a leveling off to a plateau
near 0.55.  This saturation is highly pronounced for $\Gamma = 4.5$,
5.5 and 6.5, but much less so for $\Gamma = 7.5$.  The second stage is
identified by an abrupt increase in packing fraction, which becomes
less steep and smaller as $\Gamma$ increases.  This increase in
packing fraction coincides with a nematic ordering of the material,
during which the material aligns vertically parallel to the tube
walls.  Figure~4 shows snapshots of particle configurations during
this alignment process, taken midway down the height of the tube for
$\Gamma = 4.5$: (a) and (b) are side views, as seen through the tube
wall, of the initial, randomly packed state, and the highly aligned
arrangement of the outer layer at some point toward the end of the
ordering stage, respectively.

\begin{figure}
\centerline{
\psfig{file=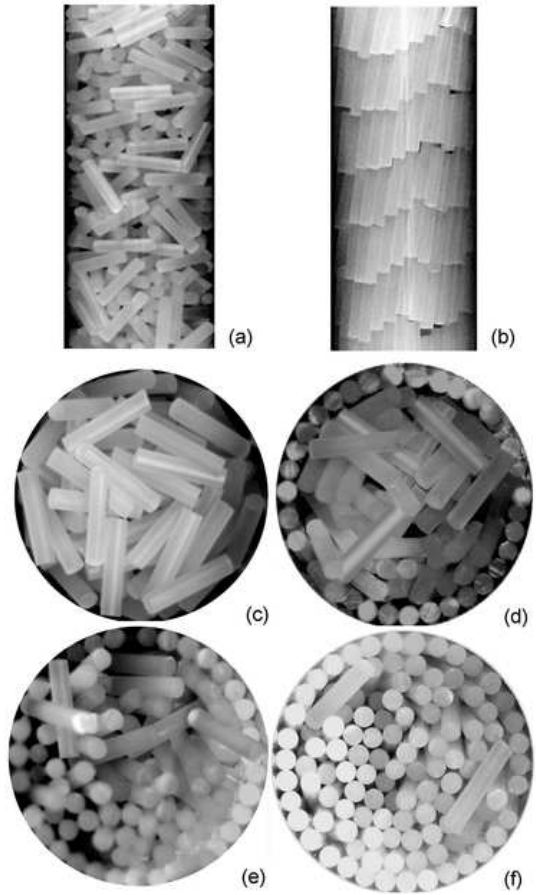,width=3in}}
\vspace{2ex}
\caption{
Snapshots of particle configurations at various times during the
packing evolution. (a) The particle configuration, as viewed from the
side, is disordered after the material is initially filled into the
column. (b) After relaxation caused by many thousands of taps the
particles line up vertically and can form a highly ordered, nematic
state. The progression from disordered to ordered is best seen by
carefully removing the material in the upper half of the column and
viewing from above (c-f).  The initial, disordered state (c), evolves
into a state with order only along the wall (d) after about 2000 taps
at $\Gamma = 4.5$ . After about 6000 taps, order has propagated in
from the walls (e), although the particles are not yet perfectly
aligned with the vertical in a nematic state. The long time behavior
(f) is a nematic state with a $\Gamma$-dependent degree of disorder in
the interior, similar to that viewed from the side in (b).  }
\label{fig:photos}
\end{figure}

Images (c)-(f) show top views of the packing interior, obtained after
careful vacuuming out the material in the upper half of the column.
The initial state~(c) shows the material as poured (but not fluffed).
After 2000 taps~(d), the particles have lined up along the edge of the
tube, while its interior remains disordered (start of ramping domain).
After 6000 taps~(e), the particles in the interior also have begun to
orient vertically.  The final, steady-state regime is a dense, highly
aligned configuration~(f). This image sequence demonstrates that the
vertical alignment starts at the tube walls and proceeds inward.  We
find that the start of the ordering stage varies slightly with height
along the tube, for $\Gamma < 5$ happening somewhat earlier at the top
of the tube and moving downward (see Fig.~3a).  At higher
accelerations, $\Gamma = 6.5$ and 7.5, the vertical alignment at the
end of the ordering stage is less perfect and at the highest
accelerations explored, $\Gamma = 7.5$, the ordering stage becomes
less distinct, particularly for the lower regions of the container.
Final, steady-state configurations at these accelerations are very
similar to those in Figure~4e, where only the outside is well aligned
while the inside still exhibits numerous defects.

In the final steady-state regime both the initial relaxation and
subsequent vertical ordering have saturated and the packing fraction
is found to fluctuate around a constant, asymptotic value (Fig.~3).
This final packing fraction typically increases with depth below the
free surface (in our data for $\Gamma= 4.5$ the differences may be too
small to be significant).  For accelerations below $\Gamma = 4.5$ we
found the dynamics to be exceedingly slow, preventing an asymptotic
state from being reached by $t = 10^5$, the longest tapping interval
explored in these experiments.  

In Fig.~5 we show the average overall packing fraction as determined
by the total height of the material in the tube, $\rho_h(t)$.  Clearly
visible is the significant increase of final packing fraction for
$\Gamma = 5.5$ and below.  In general, $\rho_h(t)$ mimics the three
stages seen in $\rho(t)$.  However, the onset of the vertical ordering
regime is not as abrupt for $\rho_h(t)$ as it is for $\rho(t)$, nor is
it preceded by the slight dip in volume fraction which ends the first
domain in Figure~3.  A direct comparison of $\rho_h(t)$ with the
height-averaged packing fraction, $<\rho(t)>_h$, obtained from the
capacitor data in Fig.~3, is shown in Fig.~6.  As we will discuss
below, the differences between the two types of measurement reflect
the fact that the active volume responsible for $\rho(t)$ includes
only a fraction of the material near the tube wall.  Also included in
Figs.~5 and 6 is a trace for $\Gamma = 4.5$ obtained under different
initial conditions:  instead of initial fluffing, the material was
merely dropped into the tube and then tapped.  We note that the memory
of the preparation conditions persists only up to about 20 taps, after
which traces for different initial conditions coincide.

\begin{figure}[t]
\centerline{
\psfig{file=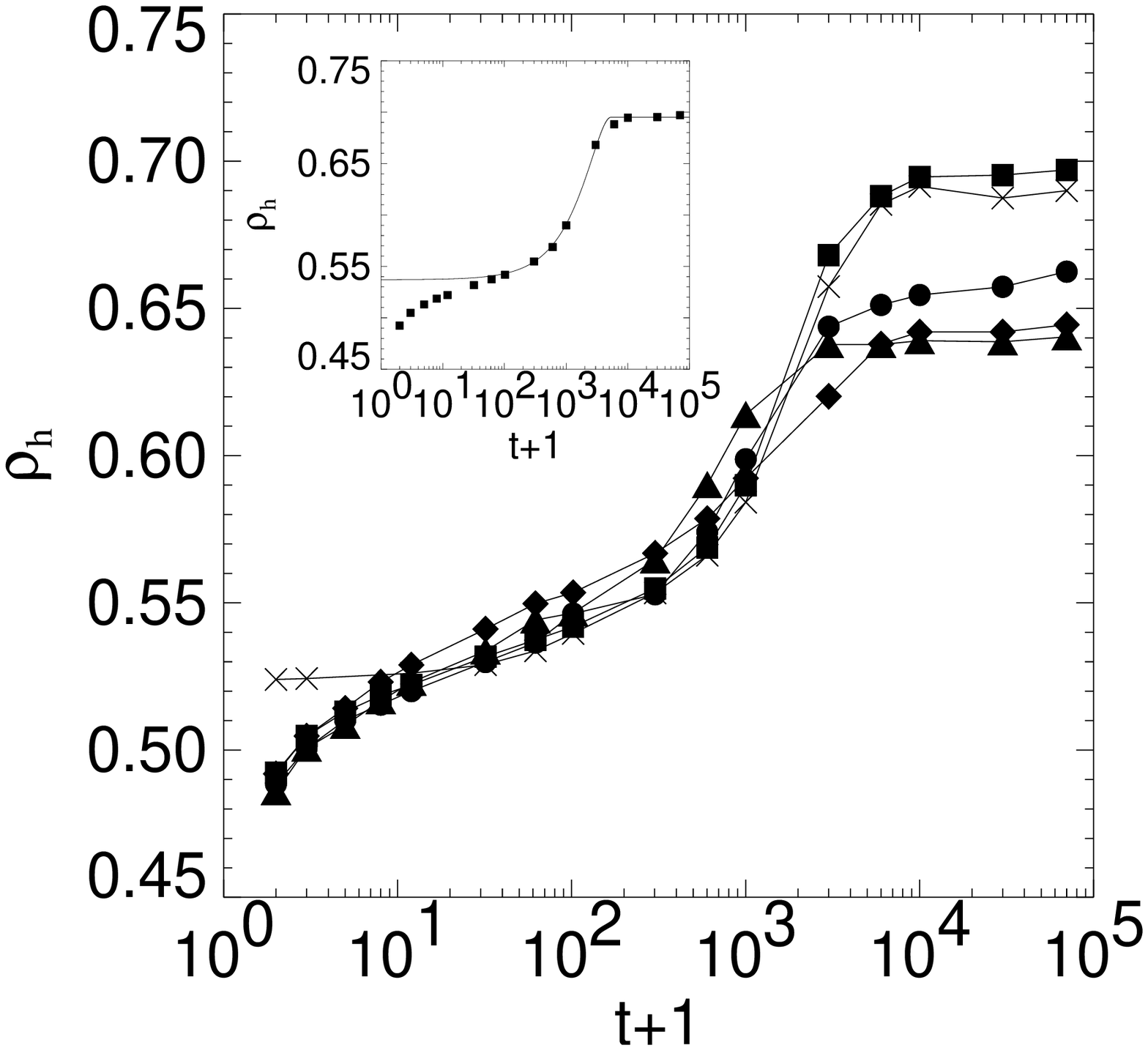,width=3in}}
\vspace{2ex}
\caption{
The time evolution of packing fraction as determined by column height,
$\rho_h(t)$, for shaking intensities $\Gamma = 4.5, 5.5$, 6.5, and
7.5, represented by squares, circles, triangles, and diamonds,
respectively. The three stages seen in the capacitance data in Fig.~3
are also visible here.  The crosses represent $\rho_h(t)$ for an
experiment with different initial conditions in which the fluffing
step in the system preparation was omitted.  The inset shows a fit of
the packing fraction for $\Gamma = 4.5$ throughout the vertical
ordering regime to${\rho_h(t) - \rho_h(t_i)
\over \rho_h(t_f) - \rho_h(t_i)} = 2{t-t_i \over t_f - t_i} - ({t-t_i
\over t_f-t_i})^2$, where $t_i$ and $t_f$ are the
beginning and ending time of the vertical ordering stage,
respectively, as discussed in Section IV.  }
\label{fig:heights}
\end{figure}

\begin{figure}[t]
\centerline{
\psfig{file=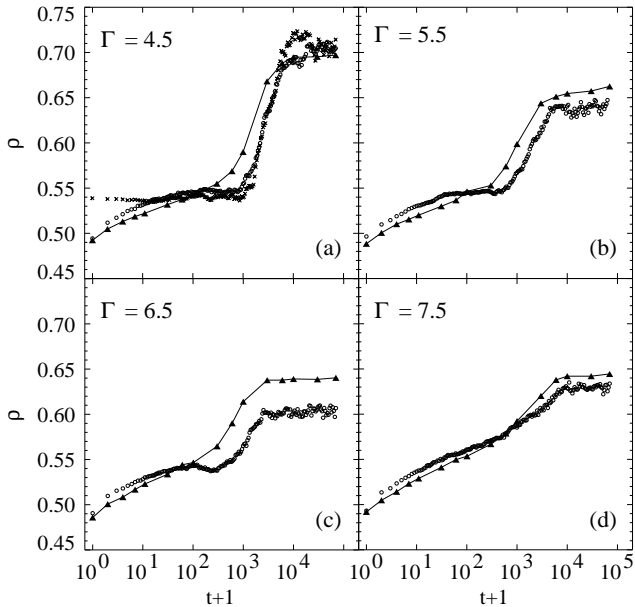,width=3.3in}}
\vspace{2ex}
\caption{
Comparison of packing fraction evolution as determined by height and
capacitance. The packing fraction as determined by height,
$\rho_h(t)$, is shown as solid triangles. The average packing fraction
of the column, $\langle\rho(t)\rangle_h$, determined by averaging the capacitive
measurements in Fig.~3, is shown as open circles. The difference
between the curves reveals the existence of a radial packing fraction
profile which evolves over time as the system orders.  The crosses
shown in (a) represent $\langle\rho(t)\rangle_h$ for different initial conditions
in which the fluffing step in the system preparation was omitted.  }
\label{fig:comparison}
\end{figure}

Results from the second type of experiment are given in Fig.~7 for
$\Delta t = 100$ (a) and 1,000 (b), using $\Delta\Gamma = 1.0$ in both
cases.  The ramp rates are fast enough that the material does not have
time to reach the steady-state during the first pass.  Consequently,
the packing fraction initially does not depend on acceleration alone,
but also strongly on the vibration history:  As $\Gamma$ is ramped
first up, and then down and up again several times, $\rho_h$ slowly cycles
toward a reversible regime  in which $\rho_h(\Gamma)$ becomes
monotonic.  For fast rates, as in Fig.~7a, a large number of cycles
may be required;  for slower ramp rates the reversible regime may be
approached much earlier (Fig.~7b).

\begin{figure}[t]
\centerline{
\psfig{file=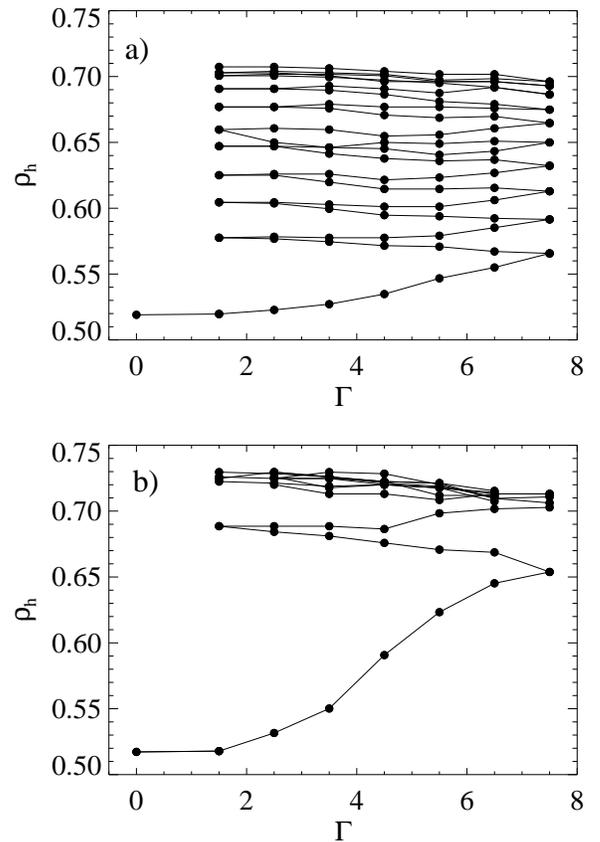,width=3in}}
\vspace{2ex}
\caption{
Packing fraction, $\rho_h$, at various stages during experiments in
which the system was not allowed to reach a steady-state before
changing the acceleration $\Gamma$.  In (a), the system was tapped for
$\Delta t = 100$ taps at each acceleration as $\Gamma$ was ramped from
1.5 to 7.5 and back for several cycles in steps of $\Delta\Gamma =
1.0$.  The system increases in packing fraction irreversibly before
reaching a reversible steady-state branch where $\rho$ becomes a
monotonic function of $\Gamma$. In (b) the time interval was increased
to $\Delta t = 1000$, allowing the system to relax more at each
acceleration value.  The reversible branch is reached after only 1.5
cycles.  }
\label{fig:cycling}
\end{figure}

\section{Discussion}
Two central observations from recent, systematic compaction
experiments~\cite{nowak97a,knight95,nowak98,nowak97b} with spheres
were a) at fixed acceleration the logarithmically slow approach in
time toward the final steady-state packing fraction, and b) the
existence of memory effects in which the final state can depend on the
initial sample preparation and on the acceleration history.  In
particular, from analysis of the packing fraction fluctuation spectra
it was found that there is an intrinsic, broad range of relaxation
time scales, many of which, however, are effectively ``frozen out'' if
the acceleration stays below certain values~\cite{nowak98,kolan99}.

Much of this behavior qualitatively carries over to the cylinder-shape
particles investigated here.  The data in Figs.~3 and 6 indicate a
non-exponential relaxation of voids in the packing during the first
stage of the packing fraction evolution.  At the highest acceleration,
when effects due to particle alignment are weakest (Figs.~3d and 6d),
we find that $\rho(t)$ increases approximately logarithmically before
it eventually saturates.  Such logarithmic relaxation arises naturally
from free volume considerations, in which the ability of particles to
find and occupy any of the remaining free space decreases
exponentially in time~\cite{nowak98,boutreux97,bennaim98}.  In its
simplest form, this scenario is independent of particle shape and,
furthermore, only based on void annihilation.  Without the additional
possibility of defect or void creation, however, the system eventually
has to jam at a final packing fraction that can only be a
monotonically increasing function of applied acceleration (since
hindrance effects are more effectively overcome at larger $\Gamma$).
Instead we find that, over the range in $\Gamma$ explored, the overall
final packing fraction decreases with increasing acceleration
(Figs.~5,~6).  As for spheres, this clearly indicates defect
production in response to tapping.  Nevertheless, as Fig.~7 shows,
memory of the tapping history is not easily destroyed and can persist
over ten-thousands of taps: if during this time the tapping intensity
is changed, the system irreversibly moves into a new state of higher
packing fraction, independent of whether the tapping intensity is
increased or decreased~\cite{josserand_note}.  Only at sufficiently
high acceleration and after sufficiently many taps is the reversible
regime reached where the packing fraction depends monotonically on
acceleration.  This situation is similar to super-heating or
super-cooling in thermal systems.

The prolate, anisotropic particle shape amplifies the ability of 3-D
cylinder packings to sustain large voids.  This is demonstrated by the
low packing fraction ($\rho(0) = 0.49$ in Fig.~3 compared to 0.59 for
spheres in the same experimental set-up~\cite{knight95}) for the
loosest, mechanically stable configuration, by the large overall
compaction range ($0.49 < \rho < 0.72$ in Fig.~3a compared to $0.59 <
\rho < 0.65$ for spheres~\cite{knight95}), and by the large
fluctuations in $\rho(t)$ in the steady-state.

As Fig.~7b shows, packing defects can be annealed out, first by either
``heating'' or ``cooling'' along an irreversible branch and, once the
reversible branch is reached, by ``cooling'' to final packing
fractions as high as 0.73.  We note that this value is roughly 10\%
smaller than the random close packing density in two
dimensions~\cite{bideau86}, $\rho_{\rm rcp}^{\rm 2D} = 0.82$ , defined
as the maximum packing fraction beyond which a transition to an
ordered triangular array would be necessary.  Along the reversible
branch, memory of the acceleration history is erased as long as
``heating'' or ``cooling'' steps are taken at sufficiently slow rates
$\Delta\Gamma/\Delta t$.  However, the maximum allowable rate itself
depends on how far the system has evolved:  With ${\Delta\Gamma/\Delta
t} = 10^{-2}$, in Fig.~7a,  the system clearly was cycled too fast
(``superheated'' as well as ``supercooled'') and stayed in the
irreversible regime even at the highest $\Gamma$ until it had time to
relax.  But, once the steady-state is reached, the same fast rate
produces very little, if any, ``supercooling''.

Qualitatively, the response to acceleration cycling seen in Fig.~7 is
similar to that obtained for spherical particles\cite{nowak97a}.
Presumably, therefore, concepts based on frustration due to volume
exclusion alone may model the observed behavior.  However, the shape
of $\rho(t)$ differs dramatically from the sphere case because of the
vertical ordering regime.  Presently available
models~\cite{edwards89,boutreux97,caglioti97,gavrilov98,head98,linz98,bennaim98,barker93,coniglio96,nicodemi97,kolan99,duke90,hong94,nicodemi00}
for $\rho(t)$ do not account for the orientational degree of freedom
and thus are not applicable without modification.

The most crucial difference between sphere and cylinder packings comes
from the tendency of cylinders to align along their long axis, both
with each other and with the container walls.  Vertical alignment
along the direction of the tube walls becomes noticeable after initial
void relaxation has started to saturate (typically after $\approx 100$
taps in Figs.~3 and 6).  After the void relaxation process saturates,
vertical particle ordering becomes the dominant mechanism for
compaction.  Note that, over the acceleration range explored, the
material once vertically aligned is highly stable against
reorientation.

This ordering process bears resemblance~\cite{mounfield94} to the
transition from isotropic to nematic phases in liquid
crystals~\cite{chaikin95} and for hard rods in thermal
systems~\cite{onsager49}.  As in a nematic, particle motion in the
aligned granular state is found to occur mainly by translation
parallel to the cylinder axis.  In cases where the aligned system was
particularly carefully cooled, we also observed vertical stacking
(seen for the outermost particle layer in Fig.~4b) akin to
smectic-type phases in liquid crystals.

As seen from Fig.~4, at any given height inside the tube, the first
particles to vertically align are those in contact with the tube
walls.  Alignment then progresses horizontally inward, until the whole
tube cross-section is ordered.  This progression is also detected
non-invasively from the difference in the packing fraction values
obtained by filling height ($\rho_h(t)$) and by capacitance
($\rho(t)$, $\langle\rho(t)\rangle_h$).  The fraction of particles near the tube
walls covered by the active measurement volume of the capacitors is
less than 50\% (Fig.~1b), making the capacitive measurements more
indicative of the packing fraction in the central tube region, along
its axis.  For this reason, $\langle\rho(t)\rangle_h$ is less than $\rho_h(t)$
during the vertical ordering stage until the ordering front has
reached the tube center and the final steady state is obtained
(Fig.~6).

In the simplest possible picture, the increase in packing fraction
during the ordering stage is solely due to conversion of disordered
three-dimensional particle arrangements with large void space to a
dense, highly aligned configuration of essentially two-dimensional
character.  A model of this process can be constructed as follows.  We
assume that cylinders in the disordered state can align vertically
only in the presence of previously aligned neighbors that act as
nucleation sites (this assumption of an essentially step-like front
separating the isotropic from the nematic regions is supported by the
calculations of Baulin and Khokhlov~\cite{baulin99}). Then the
ordering proceeds at a rate ${\partial N/\partial t} = {\rm p}n$,
where $N$ is the total number of already lined-up cylinders behind the
ordering front, $n$ is the number of open nucleation sites and p a
fixed probability for alignment.  Through the two-dimensional packing
fraction for discs, $N$ and $n$ are related to the size of the ordered
area and its inner perimeter, respectively.  For a cylindrical tube we
find, $n \propto
\sqrt{N_f - N}$  , where $N_f$
is the final number of aligned particles at $t = t_f$, the end of the
ordering stage.  The square root dependence reflects the fact that the
number of nucleation sites shrinks as the ordering front advances
radially from the tube wall toward the tube center (the tube wall acts
as a ring of nucleation sites at $t = t_i$ when $N(t_i) = 0$).  As a
result, the fraction of ordered cylinders increases with tap number as
${N(\tau)\over N_f}=2\tau - \tau^2$ for $0 \le \tau \le 1$, where
$\tau={t-t_i\over t_f -t_i}$ is the normalized tap number.  The
measured increase in packing fraction is then directly proportional to
$N(\tau)/N_f$.  If we take as the start of the ordering stage ($t =
t_i$ in the above expression) the point in time at which $\langle\rho\rangle_h$
first falls below $\rho_h$, we find that the above functional form for
$N(\tau)$ fits the data quite well (inset to Fig.~5).

Radial packing fraction gradients, detected by differences between
$\langle\rho(t)\rangle_h$ and $\rho_h(t)$, are also found in the initial
relaxation stage and in the final steady-state regime.  During void
relaxation, the capacitive data in Fig.~6 typically lie above those
from the height measurements, indicating a slightly higher packing
fraction in the central region, away from the walls.  This is a
consequence of two effects.  First, the tube wall can provide stable
pinning, preventing low-packing-fraction configurations from collapse
(initial void collapse is accelerated if the material is merely
dropped into the column, rather than carefully fluffed, as seen in
Fig.~6a).  Second, next to the tube wall, the packing fraction
naturally is reduced due to excluded volume (unless the packing
configuration is commensurable with the tube volume).  Of course, once
alignment along the tube wall has begun, the outer regions become
denser than the tube interior and the traces for $\langle\rho(t)\rangle_h$ and
$\rho_h(t)$ cross.  The slight dip present in $\rho(t)$ or
$\langle\rho(t)\rangle_h$ at the end of void relaxation and the beginning of the
ordering stages (for $\Gamma \le 6.5$) indicates that the packing
fraction temporarily decreases in the central tube region.  We
speculate that this local dilation might be required to allow
particles to rotate and align with the tube wall.

At high accelerations, $\Gamma \ge 5.5$, a radial gradient in the
packing fraction remains even in the steady-state regime (Fig.~6).  At
these accelerations, particles near the wall remain highly aligned
while the interior exhibits packing defects (similar to Figs.~4d,~e).
Consequently, the capacitively measured steady-state packing fraction
values fall below those from the height measurements (Fig.~6b-d).
Interestingly, for $\Gamma = 7.5$ this gradient develops only at large
times.  Conversely, for $\Gamma = 4.5$ the fact that $\langle\rho(t)\rangle_h$ and
$\rho_h(t)$ coincide at large times indicates a highly uniform packing
fraction profile across the tube once the asymptotic state is reached
(cf.~Fig.~4e).

As far as particle configurations are concerned, the simulations of
prolate ellipsoid packings by Buchalter and
Bradley~\cite{buchalter92,buchalter94} predict that the pouring and
packing process by itself should lead to a certain degree of nematic
ordering (``nematic glass'').  In a (infinite) system without vertical
side walls this would be a consequence of minimizing the gravitational
potential energy during particle deposition: the first rods hitting
the container bottom would tend to lay flat and thus induce horizontal
alignment of subsequent layers.  Such a state contains fewer large
voids than a completely isotropic packing configuration and is thus
denser.  In our experiments, this might be reflected in the difference
between the as poured and fluffed traces for $\Gamma = 4.5$ in Figs.~5
and 6a.  Indeed, as the side and top views in Figs.~4a and c show for
the poured initial state, there is a preference for cylinders to
orient toward the horizontal.  As the eventual merging of the
$\rho(t)$ curves for different initial conditions indicates, the same
vertically ``flattened'' packing state is also reached from the more
isotropic, fluffed initial condition, namely as most large voids have
collapsed and $\rho(t)$ starts to level off (Fig.~6a).  At least for
accelerations $\Gamma \le 6.5$ this packing state can be identified
with a packing fraction $\rho \approx 0.55$

Our present system is too small (in terms of lateral extent) to
cleanly test whether the packing configuration at the end of the void
relaxation stage corresponds to a horizontally orientated ``nematic
glass'' state.  According to Ref.~\cite{buchalter92} shaking is
expected to eventually break up this orientational order and our data
are certainly compatible with this for long times and/or high
accelerations.  However, vertical shaking in the presence of the tube
sidewalls, rather than merely reducing the degree of horizontal
ordering, provides a strong incentive for rods to line up vertically,
similar to the situation in a thermal system of rods where the loss in
rotational entropy is compensated by a gain in free volume accessible
to translations.  (As we mentioned above, this gain may be the cause
of the small dip in the packing fraction that we pick up by the
capacitive measurements  in Figs.~3a-c before the vertical ordering
changes the overall particle packing fraction.)  In principle, this
argument might also lead to some initial vertical ordering along the
tube walls from the pouring process.  From Figs.~4a and c, and also
from the fact that the initial densities $\langle\rho(0)\rangle_h$ and $\rho_h(0)$
coincide in Fig.~6, we find, however, no evidence of such alignment,
demonstrating that gravitational potential energy far outweighs
rotational entropy unless it is mitigated by the applied acceleration
during tapping.

\section{Conclusions}
We have extended our investigations of the relaxation behavior of
nonthermal, granular material to highly anisotropic, cylindrical
particle shapes.  Using a combination of non-invasive, capacitive
probes and video imaging, we have traced both the local and global
packing fraction $\rho$ as well as the evolution of the packing
configurations (Fig.~4) from an initial, low-packing-fraction to a
final, high-packing-fraction state under applied mechanical
excitations.

We observe many qualitative features also seen in the relaxation of
sphere packings and find them amplified by the particles' anisotropy.
This includes an even wider range of metastable configurations and
thus a larger span of accessible packing fractions (Fig.~3), as well
as memory effects in the irreversible branch of $\rho(\Gamma)$ that
are much more pronounced (Fig.~7).

In contrast to sphere packings, which tend to end up in disordered
configurations even after prolonged tapping and cycling, we find clear
evidence that particle anisotropy can drive ordering.  This is most
strikingly observed in $\rho(t)$ where we can distinguish three
characteristic stages (Figs.~3 and 6):  First, a void relaxation stage
takes the initially isotropic arrangement to an intermediate packing
fraction near $\rho = 0.55$.  This is a state of vertically collapsed
and thus predominantly horizontally oriented particle configurations
which may correspond to the nematic glass state seen in computer
simulations~\cite{buchalter94}.  This state, however, turns out to be
unstable to continued vertical excitations.  During a second stage,
the vertical ordering regime, particles re-align their long axes
vertically.  This ordering process starts from the container walls,
moves inward and eventually leads to an ordered, nematic-type
configuration.  We have shown that a simple model can account for the
change in packing fraction during this ordering process.  The imposed
boundary conditions at the wall stabilize the nematic state,
effectively playing a role similar to strong flow
fields~\cite{mounfield94} or packing fraction
gradients~\cite{baulin99}.  The third, and final, stage is a
steady-state with large fluctuations around an average packing
fraction set by competing defect annihilation and creation within the
nematic-type particle arrangement.

These results provide a first experimental step toward the full
exploration of the effect of anisotropy on ordering in strongly
non-equilibrium, nonthermal systems.  Our experiments used a fixed
aspect ration of close to four.  Larger aspect ratios as well as less
rigid particles are expected to hinder the transition from isotropic
to nematic configurations, and there is also evidence that oblate
particles should order differently from the prolate ones investigated
here~\cite{buchalter94}. Another intriguing extension concerns
mixtures of spheres and rods.  For a thermal system of this type,
novel micro-phase separated patterns, such as lamellae of spheres and
rods, have recently been found experimentally~\cite{adams98}, while
theoretical models for the granular equivalent would suggest
macroscopic phase separation~\cite{edwards94}.  Finally, for none of
the non-spherical systems has the spectrum of fluctuations around the
steady-state been explored yet.

\section*{Acknowledgments}
We wish to thank Damien Dawson, Allan Smith, Tom Witten and, in
particular, Sidney Nagel for many helpful and illuminating
discussions.  This work was supported by the NSF under Award
CTS-9710991 and by the MRSEC Program of the NSF under Award
DMR-9808595.

\end{document}